\documentclass{article} 

\usepackage{epsfig}
\usepackage{amsmath}
\usepackage{cite}
\usepackage{a4wide}

\newcommand{\be}{\begin{equation}}
\newcommand{\ee}{\end{equation}}
\newcommand{\bea}{\begin{eqnarray}}
\newcommand{\eea}{\end{eqnarray}}
\newcommand{\bm}[1]{\mbox{\boldmath $#1$}}
\newcommand{\abs}[1]{\left| #1 \right|}

\title{Nonequilibrium Statistical Mechanics of Swarms of Driven Particles}

\author{Werner Ebeling\footnote{ebeling@physik.hu-berlin.de},
  Udo Erdmann\footnote{udo.erdmann@physik.hu-berlin.de}\\
  {\normalsize Institute of Physics, Humboldt-University,}\\
  {\normalsize Newtonstra{\ss}e 15 112489 Berlin, Germany} }

\date{}
\begin{document}
\maketitle

\begin{abstract}
  As a rough model for the collective motions of cells and organisms we
  develop here the statistical mechanics of swarms of self-propelled
  particles. Our approach is closely related to the recently developed theory
  of active Brownian motion and the theory of canonical-dissipative systems.
  Free motion and motion of a swarms confined in an external field is studied.
  Briefly the case of particles confined on a ring and interacting by
  repulsive forces is studied. In more detail we investigate self-confinement
  by Morse-type attracting forces. We begin with pairs $N = 2$; the attractors
  and distribution functions are discussed, then the case $N > 2$ is
  discussed. Simulations for several dynamical modes of swarms of active
  Brownian particles interacting by Morse forces are presented. In particular
  we study rotations, drift, fluctuations of shape and cluster formation.
\end{abstract} 

\section{Introduction}

We consider here the collective modes and the distribution functions of finite
systems of free particles, of particles confined in an external field and
systems of interacting Brownian particles including active friction. This is
considered as a rough model for the collective motion of swarms of cells and
organisms \cite{ViCzBeCoSh95,DeVi95,ShiSuMiHaSa96,CziVi00}.

First we will discuss the basic model of active Brownian particles and
introduce several models of active friction. The energy supply is modeled by a
velocity-dependent function.  Beside the classical model of negative friction
due to Rayleigh \cite{Ra45,Kl95,MiZa99} we will discuss the model of
Schienbein and Gruler \cite{SchiGr93} and the so-called depot model which has
been derived from concrete assumptions about the energy supply
\cite{SchwEbTi98,EbSchwTi99,ErEbSchiSchw99,MaEbVe99,EbErDuJe99}. We include
into the model also a weak global coupling to the velocity of the center of
mass.

Then the motion of a swarm of noninteracting particles which is confined in an
external field is studied. Furthermore the case of particles confined on a
ring and interacting by repulsive forces is investigated. Next we will
investigate self-confinement by Morse-type attracting forces. The Morse
potential is defined by
\begin{equation}
  U = \frac{A}{2b}\left\{\left[e^{-b(r-\sigma)}-1\right]^2-1\right\},
\end{equation}
where $r$ is the distance of two interacting particles.  This simple model of
interactions describes repulsive and attractive interactions similar to the
Lennard-Jones potential. The Morse potential which has the right physical
shape is however much more useful with respect to analytical treatment. This
is well-known from quantum mechanics, where even exact solutions of the
Schr\"odinger equation for the Morse potential are known. Since we are very
much interested in analytical solutions we decided to use the Morse potential
also as an appropriate model for the description of the essentially
non-physical interactions in swarms.

We begin with the case of pairs $N = 2$ with Morse interactions, the
attractors of motion and the distribution functions are discussed in detail.
New solutions for the dynamical modes are presented in the deterministic and
in the stochastic description. The border of stability of the modes is
discussed. Finally the investigation is extended to systems with a large
number of active Brownian particles interacting by Morse forces. In particular
we will study the left/right rotations of pairs, clusters and swarms. We will
show that the collective motion of large clusters of driven Brownian particles
reminds very much the typical modes of collective motions in swarms of living
entities.

In the theoretical part we will make use of methods developed in the context
of the theory of the so-called canonical dissipative systems
\cite{Gr73,Gr81,Ha73,HoRy78,Eb00}. This theory is in close relation to our
recently developed theory of active Brownian particles
\cite{SchwEbTi98,EbSchwTi99,TiSchwEb99,ErEbSchiSchw99,SchwEbTi01}.

\section{Dissipative forces and equations of motion}

Let us consider two-dimensonal systems of $N$ point masses $m$ with the
numbers $1,2,...,i,...N$. We assume that the masses $m$ are connected by pair
interactions.  If the distance between the mass $i$ and the mass $j$ is
denoted by $\bm{r}_{ij}= \bm{r}_i - \bm{r}_j$, the force is $- U'(\bm{r})$.

The dynamics of the system is given by the following equations of motion
\begin{equation}
  \label{langev-or}
    \dot{\bm{r}}_i= \bm{v}_i\;,\quad 
    m\dot{\bm{v}}_i  +    
    \sum_j U'(\bm{r}_{ij}) \frac{\bm{r}_{ij}}{\abs{r_{ij}}}  
    = \bm{F}_i(\bm v_i) + \sigma^2 (\bm v_i - V) + \sqrt{2D} \bm{\xi}_i(t) 
\end{equation}
Here the dissipative forces are expressed in the form
\begin{equation}
  \bm{F}_i(\bm{v}_i) = -m \gamma(\bm{v}_i^2) \bm{v}_i
\end{equation}
The function $\gamma$ denotes a velocity-dependent friction, which possibly
has a negative part. The second term on the r.h.s. models a weak global
coupling to the average velocity of the swarm
\begin{equation}
  \bm{V} = \frac{1}{M} \sum_i m_i \bm{v}_i
\end{equation}
This way the dynamics of our Brownian particles is determined by Langevin
equations with dissipative contributions. The Langevin equations contain as
usually a stochastic force with strength $D$ and a $\delta$-correlated time
dependence.
\begin{equation}
  \label{stoch}
  \langle \bm{\xi}_i (t) \rangle=0 \,;\,\,
  \langle \bm{\xi}_i(t)\bm{\xi}_j (t')\rangle=
  \delta(t-t') \delta_{ij}
\end{equation}
In the case of thermal equilibrium systems we have
$\gamma(\bm{v})=\gamma_{0}={\rm const.}$. In the general case where the
friction is velocity dependent we will assume that the friction is
monotonically increasing with the velocity and converges to $\gamma_0$ at
large velocities. In the following we will use the following ansatz based on
the depot model for the energy supply
\cite{SchwEbTi98,EbSchwTi99,EbErSchiSchw99}
\begin{equation}
  \gamma(\bm{v}^2)=  \left(\gamma_0 - \frac{d q}{c + d v^2}\right)
\end{equation}
where $c,d,q$ are certain positive constants characterizing the energy flows
from the depot to the particle \cite{EbSchwTi99,ErEbSchiSchw99}. These
assumptions lead to the dissipative force law
\begin{equation}
  \bm{F}  = m \bm{v} \left(\frac{d q}{c + d v^2} - \gamma_0 \right)
\end{equation}
Dependent on the parameters $\gamma_0$, $c$, $d$, and $q$ the dissipative
force function may have one zero at $\bm{v} = 0$ or two more zeros with
\begin{equation}
  \label{v-0}
  \bm{v}_0^2=\frac{c}{d}\,\zeta\;,\quad \text{where}\quad
  \zeta = \frac{qd}{c \gamma_0} -1
\end{equation}
is a bifurcation parameter. In the case $\zeta > 0$ a finite characteristic
velocity $v_0$ exists. Then we speak about active particles.  We see that for
$|\bm{v}| < v_{0}$, i.e. in the range of small velocities the dissipative
force is positive, i.e. the particle is provided with additional free energy.
Hence, slow particles are accelerated, while the motion of fast particles is
damped.  In the case of small values of the parameter $\zeta$ we get the
well-known law of Rayleigh (see \cite{ErEbSchiSchw99})
\begin{equation}
  \bm{F} = m \gamma_0 \left(\zeta  - \frac{d}{c}\, \bm{v}^2\right) \bm{v}
\end{equation}
and in the opposite case of large values of $\zeta$ we get the law derived by
Schienbein and Gruler from experiments for the dynamics of cells
\cite{SchiGr93}
\begin{equation}
  \bm{F} =  m \gamma_0 \left( \frac{v_0}{|\bm{v}|} - 1 \right) \bm{v}
\end{equation} 
This way, we have shown that our depot model covers several interesting
limiting cases discussed earlier in the literature \cite{ErEbSchiSchw99}.

\section{Brownian particles in parabolic confinement}

Let us consider a finite many-particle system consisting of $N$ particles
which are driven by dissipative forces and are confined by external forces.
Interactions and global coupling are neglected so far. It is important to
understand first the motion of non-interacting Brownian particles before going
to interacting systems. For the Langevin equation of motion we postulate
\begin{equation}
  m \dot{\bm{v}}_i + \nabla U = 
  \bm F_i(\bm v_i)   + \sqrt{2 D}\, \bm{\xi}_i (t)\,.
\end{equation}
Here $U$ is the potential of the external forces and $\bm{F}_i$ is the
velocity-dependent dissipative force discussed above. Sometimes we will use
later units with $m=1$. We begin with the case of a parabolic external
potential
\begin{equation}
  U = U(\bm r) = \frac{m}{2} \omega_0^2 \bm r^2
\end{equation}
For non-interacting particles the $N-$particle distribution factorizes and it
is sufficient to treat the one-particle problem. The one-particle distribution
function is at the same time the relative density of a swarm of
non-interacting particles moving in the external field.  The Fokker-Planck
equation and the standard methods to obtain the stationary distribution
functions are in detail discussed by Klimontovich \cite{Kl95} for the
one-dimensional case. These methods were extended to the case of
two-dimensional oscillators in \cite{ErEbSchiSchw99}. In the mentioned work
however only approximative solutions could be presented. Here we make use of
the fact that our system is of canonical-dissipative type and will present a
new solution which is exact at small noise \cite{Eb01}. Let us first consider
the deterministic motion.  Under stationary conditions the particles have to
obey the requirement of balance between centrifugal and attracting forces
$v^2/r = \abs{U'(r)}$. For the harmonic potential this leads to the stationary
radius $r_0 = v_0 / \omega_0$. Actually the particles are moving in the
neighborhood of two limit cycle orbits which have the projections given above
and are located on two surfaces in the four-dimensional space corresponding to
the angular momenta $L = \pm\, v_0^2 / \omega_0$.  This way the probability is
concentrated on two closed curves in the four-dimensional phase space which
are similar to tires \cite{ErEbSchiSchw99}. In order to find the solution we
go step by step and use several relations valid on the limit cycle orbit. For
example we have $\abs L = H / \omega_0$. Further the potential and the kinetic
energy are exactly equal
\begin{equation}
  \bm{v}^2 = \frac Hm = \frac{1}{2}\, \bm{v}^2 + \frac{1}{2}\, 
  \omega_0^2 \bm{r}^2
\end{equation} 
Using this relation we may replace $\gamma(v^2)$ by $\gamma(H/m)$,
this is exact on the l.c. and a good approximation near to it. This way we 
obtain a solution for the probability density which is concentrated around 
$H \simeq m v_0^2$. 
\begin{eqnarray}
  f_1 (x_1, x_2, v_1, v_2) &=& C \left[1 + \frac{d}{2c} H\right]^{\frac{q}{2D}}
  \exp{\left[ - \frac{H}{2 kT} \right]}
\end{eqnarray}
This expression however is not yet the correct solution since the
corresponding angular momentum is not perpendicular to the coordinate plane.
In order to concentrate the probability of the two tires which are observed in
the simulations \cite{ErEbSchiSchw99} we make use of the relation $H = \pm L
\omega_0$ also valid on the l.c..  Here
\begin{equation}
  L = m \left(x_1 v_2 - x_2 v_1\right)
\label{eq:ang_momentum}
\end{equation}
is the angular momentum. On the limit cycles it has the values $L = \pm mr_0
v_0$. In the stochastic case the angular momentum is distributed around $L
\simeq \pm H / \omega_0$. We assume that this distribution is Gaussian.
Combining now both parts of the distribution in such a way that the
dissipative probability flow disappears, we obtain the following approximation
for the stationary distribution function
\begin{eqnarray}
  f_0 (x_1, x_2, v_1, v_2) &=& f_1
  \left\{ \exp\left[-\alpha_1 (H - \omega_0 L)\right] 
  + \exp\left[-\alpha_1 (H + \omega_0 L)\right] \right\}
\end{eqnarray}
In the case $D \rightarrow 0$ this approximative solution is concentrated
around the two limit cycles. From the condition that the mean energy remains
unshifted we get
\begin{equation}
  \alpha_1 = \frac{2 q d^2}{c^2}\, v_0^2
\end{equation}
Summarizing this results we may conclude that the shape of the probability in
the four-dimensional phase space is well understood from theory and numerical
simulations. However the available approximate solutions reflect only the
limit of small noise so far. Exact solutions for finite noise level are not
yet known.

Before we proceed to the general case of interacting particles let us make a
study of a limiting case of interacting systems which may be treated in an
elementary way. In the case of very strong driving, i.e. $v^2_0 \gg kT$ all
Brownian particles will move on large orbits $r_0 = v_0 / \omega_0$. This way
the particles form a ring. For more general confining forces with radial
symmetry $U(r)$ the radius $r_0$ is determined by the condition of equilibrium
between centrifugal and centripetal forces
\begin{equation}
  \frac{v_0^2}{r_0} = \abs{U'(r_0)}
\end{equation}
The effect of weak noise will be an equal distribution on the ring and some
small dispersion perpendicular to it. Switching on arbitrary repulsive forces
between the Brownian particles, e.g. an exponential repulsion (Toda forces)
between neighboring particles leads to a kind of lattice on the ring with the
average distance
\begin{equation}
  l = \frac{2 \pi r_0}{N}
\end{equation}
This way we come to the interesting conclusion that in the limit of strong
driving, confined particles with repulsive interactions will always form
rotating ring lattices.

A more detailed study of the case of Toda rings 
has been given in our earlier work \cite{EbErDuJe99,MaEbVe99,EbLaUs01}. 
Replacing the repelling
exponential Toda forces by Morse forces which have an attracting tail 
beside rotating ring lattices also clustering phenomena 
on rings are observed
\cite{DuEbEr01,DuEbErMa01}.

\section{Dynamics of self-confined clusters of driven particles}

The study of the many-body dynamics of interacting driven particles is an
extremely difficult task. Therefore we begin our investigation with the study
of 2 driven Brownian particles \cite{ErEbAn00} which are self-confined by
attracting forces. In this relatively simple case we observe already several
basic features of the dynamics of swarms. Let us consider two Brownian
particles which are pairwise bound by a radial pair potential $U(r_1 - r_2)$,
as a concrete example we may consider the parabolic case $U = (a/2)(\bm{r}_1 -
\bm{r}_2)^2$. Another example is the Morse potential defined in the
introduction. The pair of particles will form dumb-bell like configurations.
Then the motion consists of two independent parts: The free motion of the
center of mass $\bm{R} = (\bm{r}_1 + \bm{r}_2)/2$ and the mass velocity
$\bm{V} = ( \bm{v}_1 + \bm{v}_2)/2$. The corresponding coordinates are $X_1 =
(x_{11} + x_{21})/2$ and $X_2 = (x_{12} + x_{22})/2$. The relative motion
under the influence of the forces is described by the relative radius vectors
$\bm{r} = (\bm{r}_1 - \bm{r}_2)$ and the relative velocity $\bm{v} = (
\bm{v}_1 - \bm{v}_2)$. The relative coordinates are $x_1 = (x_{11}-x_{12})$
and $x_2 = (x_{12} - x_{21})$.  Let us first study the deterministic
equations. The motion of the center of mass is described by the equations
 \begin{equation}
  m\dot{\bm{V}} = \frac{1}{2} \left[\bm{F}\left(\bm{V} +\frac{\bm{v}}{2}\right)
    +\bm{F}\left(\bm{V} -\frac{\bm{v}}{2} \right) \right]
\end{equation}
The relative motion is described by 
\begin{equation}
  m\dot{\bm{v}} + U'(r) \frac{\bm{r}}{r} = \frac{1}{2} \left[\bm{F}
    \left(\bm{V} +\bm{v}\right) -\bm{F}\left(\bm{V} -\bm{v}\right) \right]
    + \sigma^2 \bm{v}
\end{equation}
This system possesses two types of attractors. The first one corresponds to a
translational motion were the two particles move nearly parallel and we have
$v^2 \ll V^2 \simeq v_0^ 2$. With this assumption we find in quadratic
approximation in $v$:
\begin{equation}
  \dot{\bm{V}} = - \left[\gamma(V^2)  + \gamma'(V^2) v^2 \right]\bm{V}
  - 2 \gamma'(V^2)(\bm{V}\cdot\bm{v})\bm{v}+ \ldots   
\end{equation}
Assuming that the term of second order ${\cal O}(v^2)$ is bounded and remains
small we may conclude from this dynamical equation that all velocity states
$V^2 \simeq v_0^2$ will converge to the attractor state $V^2 = v_0^2$. Since
$\gamma'(v_0^2) > 0$ the coupling to the relative motion may lead to an
enhancement of the translation due to an energy flow from the relative to the
translational mode. For the relative velocities we get in linear approximation
\begin{equation}
 m \dot{\bm{v}} + U'(r) \frac{\bm{r}}{r} = 
  - \bm{\Gamma} \cdot \bm{v} + {\cal O}(v^2)
\end{equation}
Here $\bm{\Gamma}$ is a tensor which we call friction tensor
\begin{equation}
  \bm{\Gamma} = \left[\left(\frac{\sigma^2}{m} + \gamma(V^2)\right) \bm{\delta} + 
    2 \gamma' (V^2) (\bm{V }\bm{V}) \right]
\end{equation}
From the dynamical equation for $\bm{v}$ we find for the relative energy
\begin{equation}
  \frac{d}{dt}\left( \frac{m}{2}\,\bm{v}^2 + U(r) \right) = 
  - \bm{v}\cdot {\bm\Gamma} \cdot \bm{v} + {\cal O}(v^2)  
\end{equation}
For translational velocities near to the root i.e.  $V^2 \simeq v_0^2$ the
tensor $\bm{\Gamma}$ is positive, having only positive eigenvalues. In this
case the r.h.s. of the energy equation is negative i.e. the energy tends to
zero, both particles collapse to the minimum of the potential. In other words
the attractor of motion is 
\begin{eqnarray}
  \bm{V} = v_0 \bm{n}\; ; \quad\bm{R}(t) = v_0 \bm{n} t + \bm{R}(0)
\end{eqnarray}
In the attractor state $V^2=v_0^2$ itself the (positive) tensor reads 
\begin{equation}
  \bm{\Gamma} = \frac{\sigma^2}{m} + 2 \gamma' (v_0^2) (\bm{V} \bm{V}) 
\end{equation}
This means, excitations with $\bm{v}$ perpendicular to $\bm{V}$ (i.e.
$(\bm{V}\bm{V}) =0$) are only weakly damped and excitations with $\bm{v}$
parallel to $\bm{V}$ show a stronger damping. With decreasing ${V^2} <
{v_0^2}$ the tensor $\bm{\Gamma}$ shows a bifurcation. This happens, when the
first eigenvalue crosses zero, at this point the translation mode is getting
unstable. As a consequence the terms ${\cal O}(v^2)$ may increase unboundedly
and the translational mode breaks down. A similar bifurcation has been found
for the one-dimensional case in \cite{MiZa99}.

Let us consider now the influence of noise, naturally we expect some
distribution around the attractors. 
In the stable translational regime we find in some approximation 
the stationary distribution
\begin{equation}
  f^{(0)}(\bm{V},\bm{v},\bm{r}) = 
  C \left(1 + \frac{d}{c} \bm{V}^2\right)^{\frac{q}{2D}} 
  \exp\left[ -\frac{1} {kT} \, 
    \left( m \bm{V}^2 +
\frac{m}{2}\bm{v}^2  + U(r) \right)\right]
\end{equation}
This corresponds to a driven motion of a free particle located in the center
of mass supplemented by a small oscillatory relative motion against the center
of mass. The solutions for the rotational model are similar to what we have
found for the case of external fields. The probability is distributed around
two limit cycles corresponding to left or right rotations.

Summarizing our finding we may state: For two interacting active particles
there exists a translational mode in which the center of mass of the dumb-bell
makes a driven Brownian motion similar to a free motion of the center of mass.
In the rotational mode the center of the dumb-bell is at rest and the system is
driven to rotate around the center of mass.  In this mode only the internal
degrees of freedom are excited and we observe driven rotations. The attractor
region of the translational state has been described above, the attractor
region of the rotational state and the exact distribution
function has still to be explored.

\section{The Dynamics of self-confined swarms of active Brownian particles}

\begin{figure}[htbp]
  \begin{center}
    \begin{minipage}{13.8cm}
      \begin{center}
        \epsfig{file=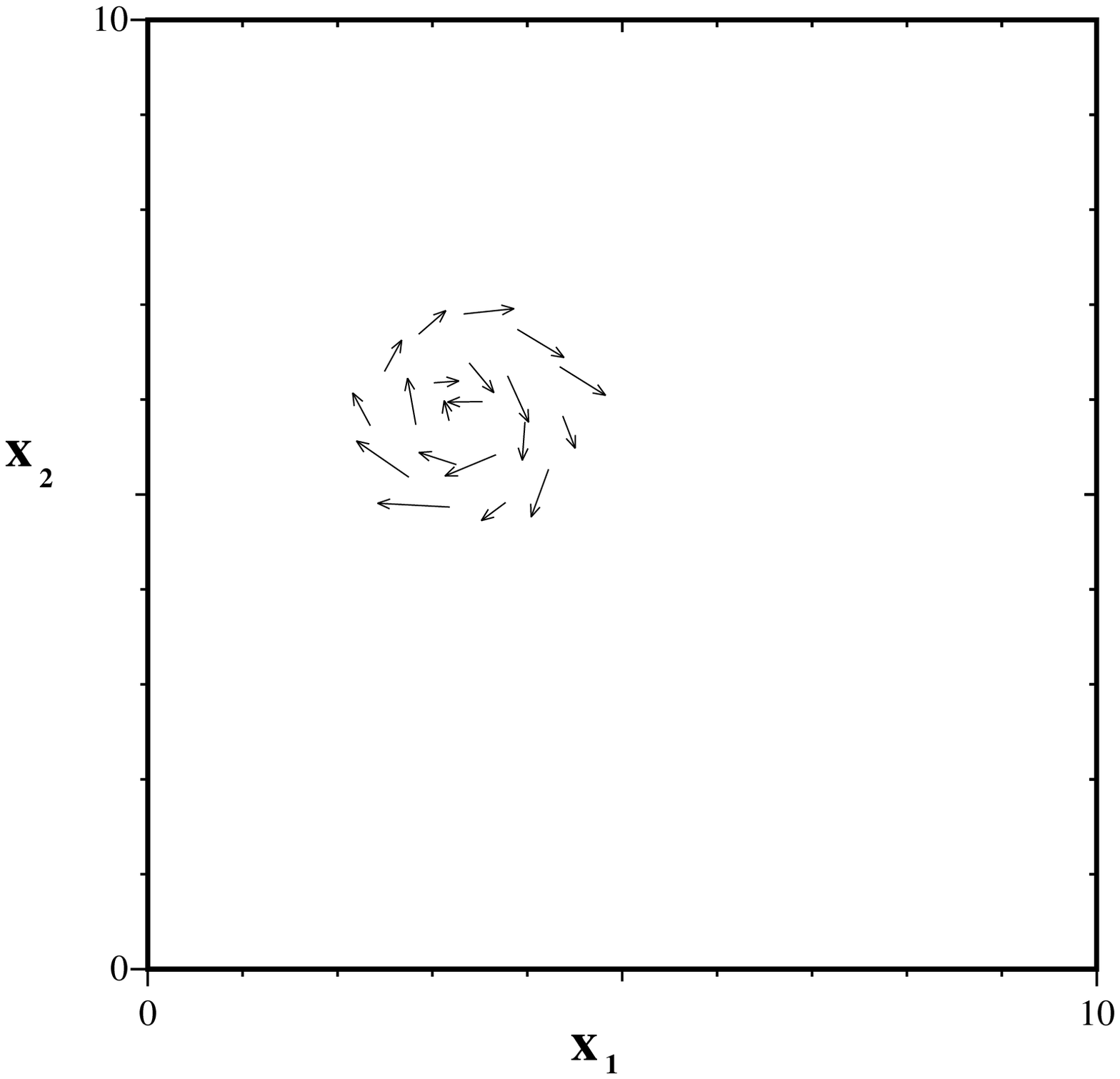, width=6.8cm}
        \epsfig{file=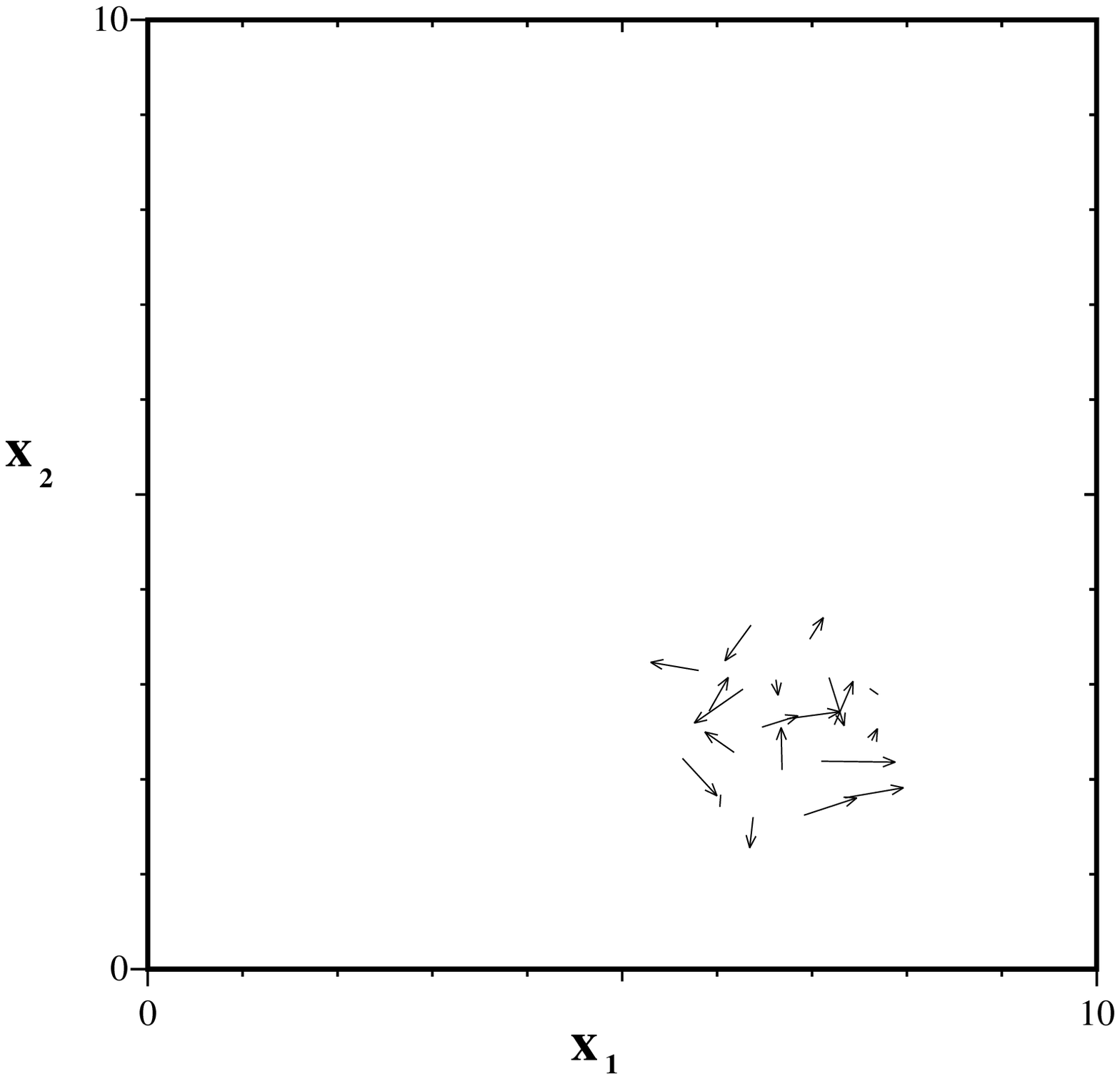, width=6.8cm}
        \epsfig{file=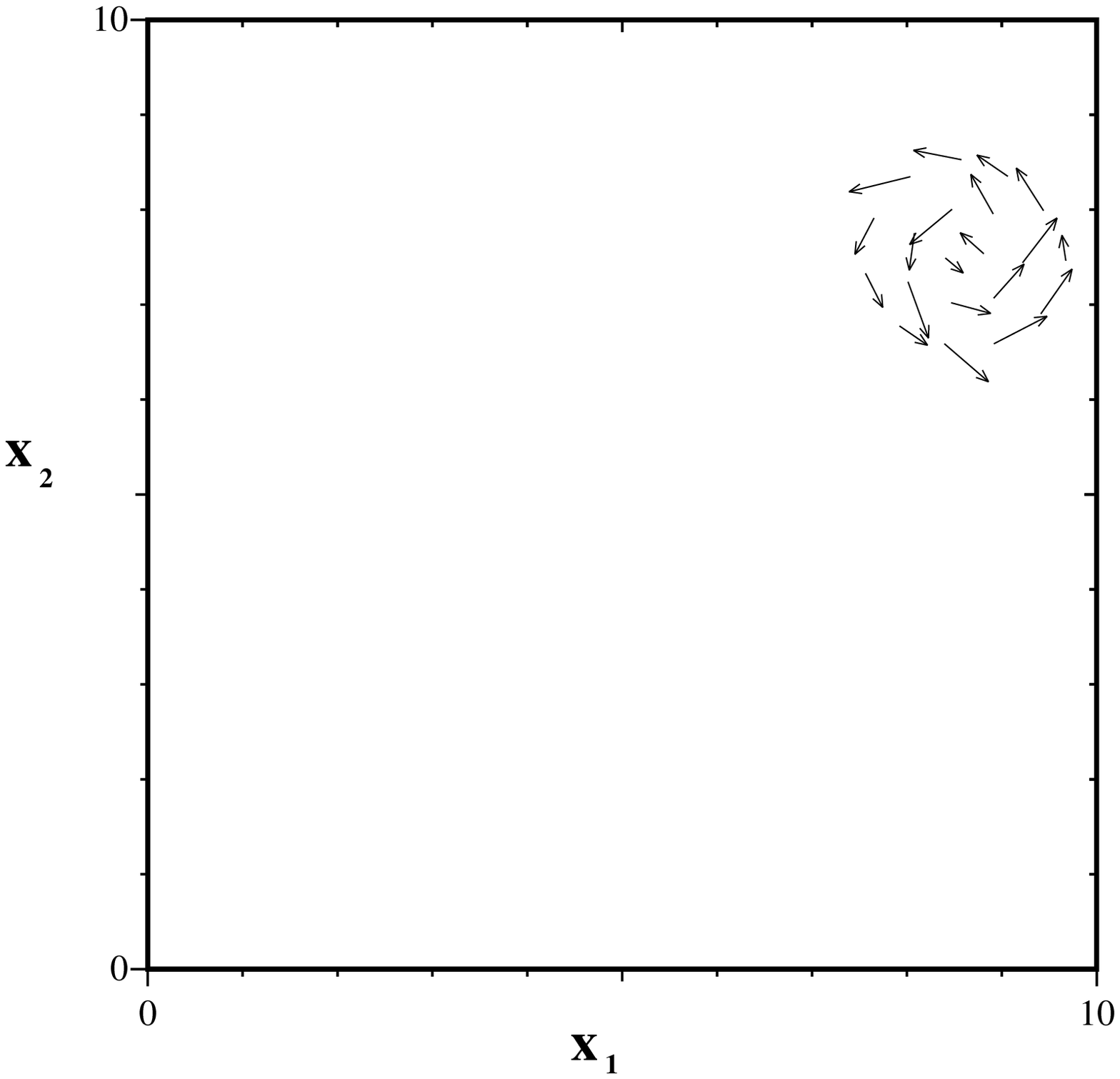, width=6.8cm}
      \end{center}
    \end{minipage}
    \vspace{0.5cm}
    \caption{Rotating cluster of 20 particles for different time steps. The
      arrows correspond to the velocity of the single particle. Because of the
      influence of noise the cluster changes the direction of rotation
      randomly from \cite{ErEbAn00}.}
    \label{fig:rot}
  \end{center}
\end{figure}

\begin{figure}[htbp]
  \begin{minipage}[t]{14cm}
    \begin{center}\noindent
      \epsfig{file=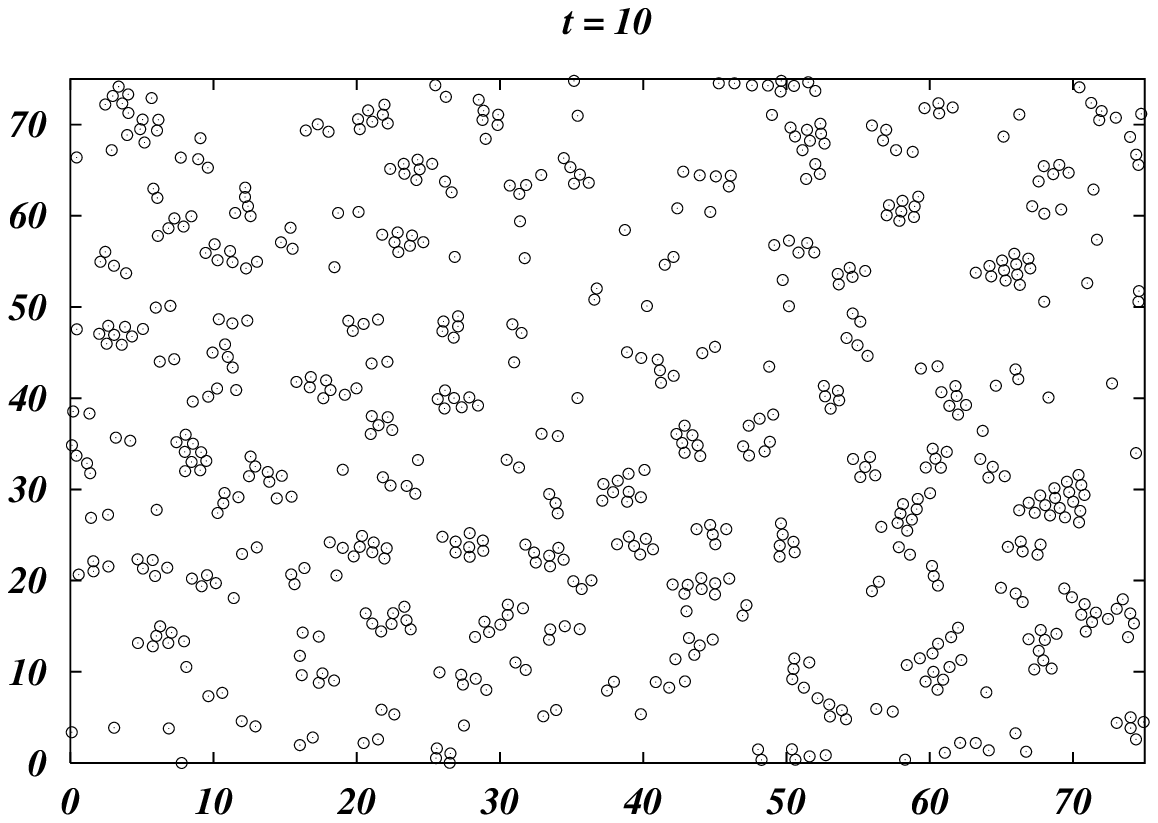,width=6.8cm}
      \epsfig{file=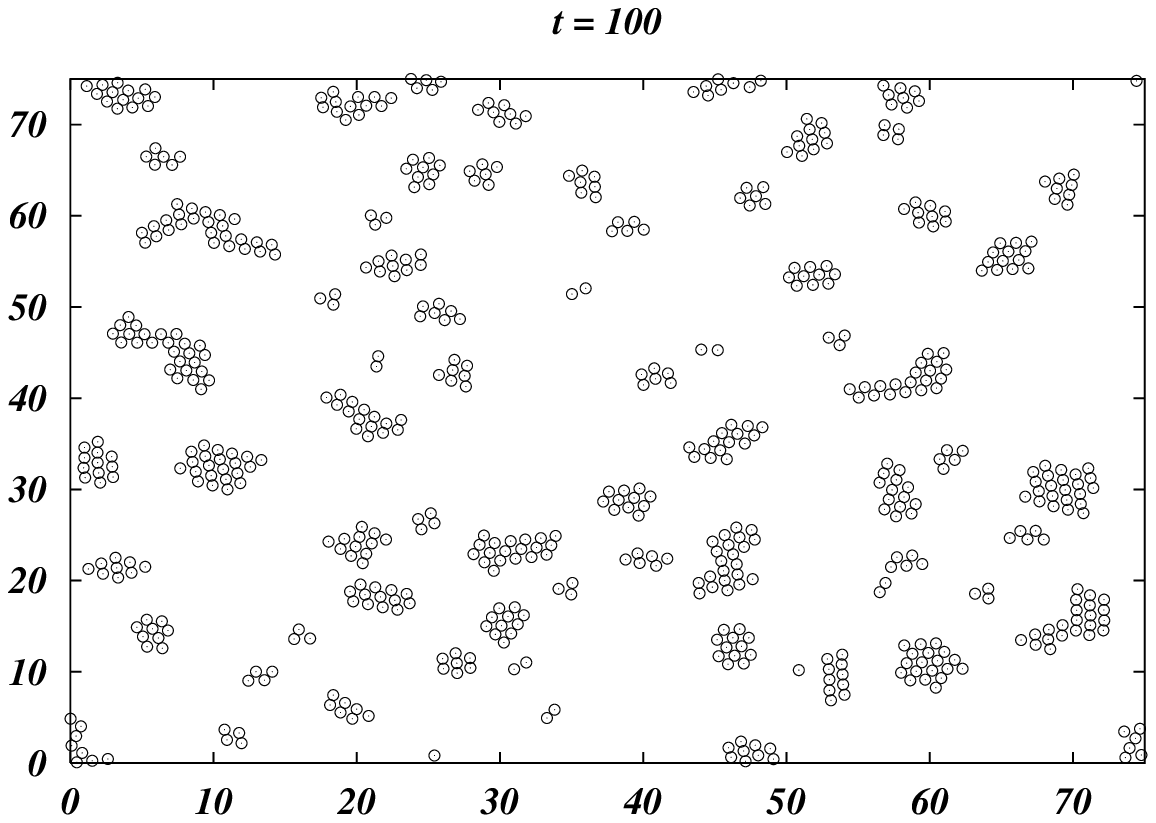,width=6.8cm}\\\vspace*{0.3cm}
      \epsfig{file=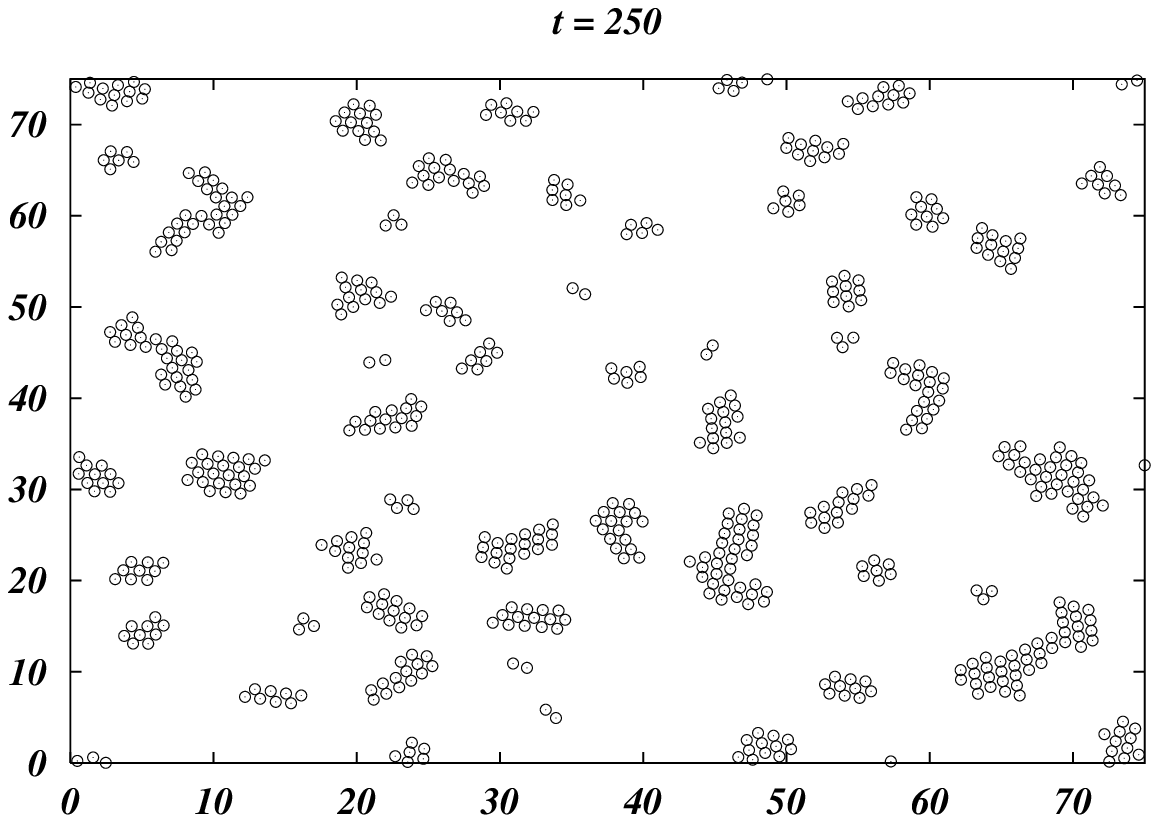,width=6.8cm}
      \epsfig{file=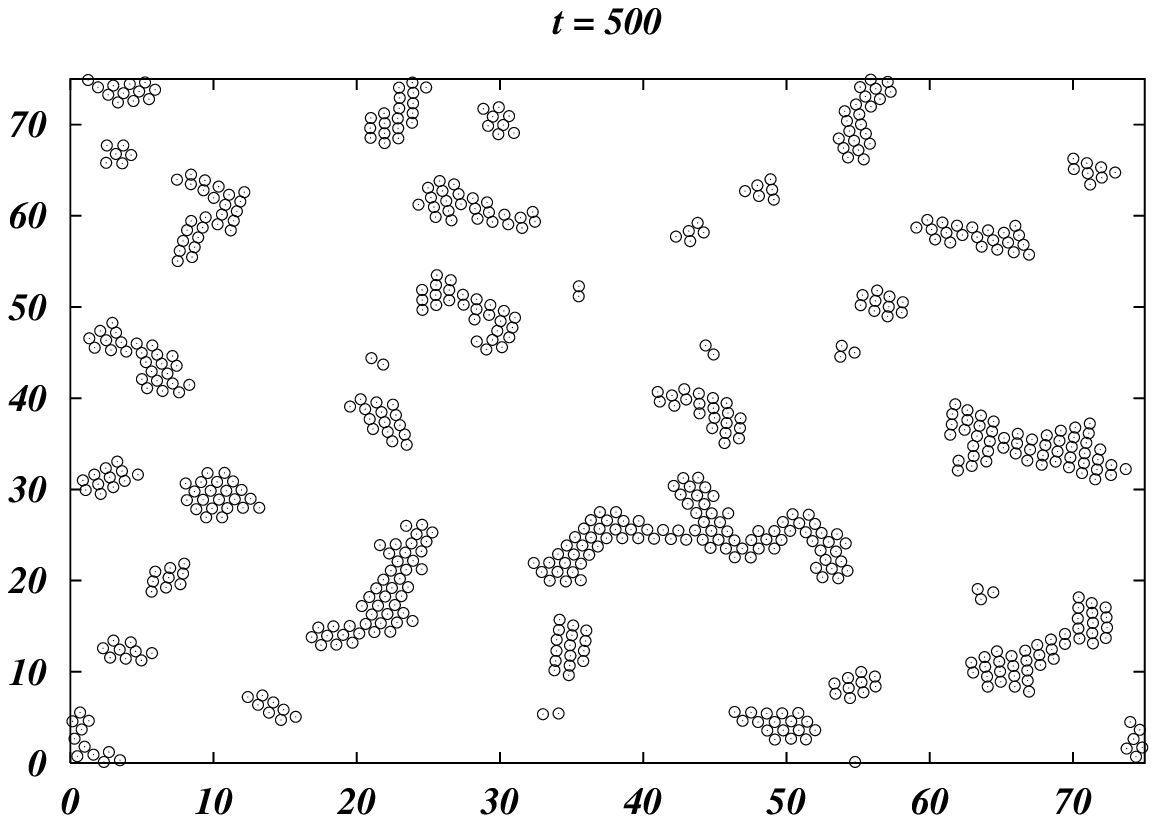,width=6.8cm}\\\vspace*{0.3cm}
      \epsfig{file=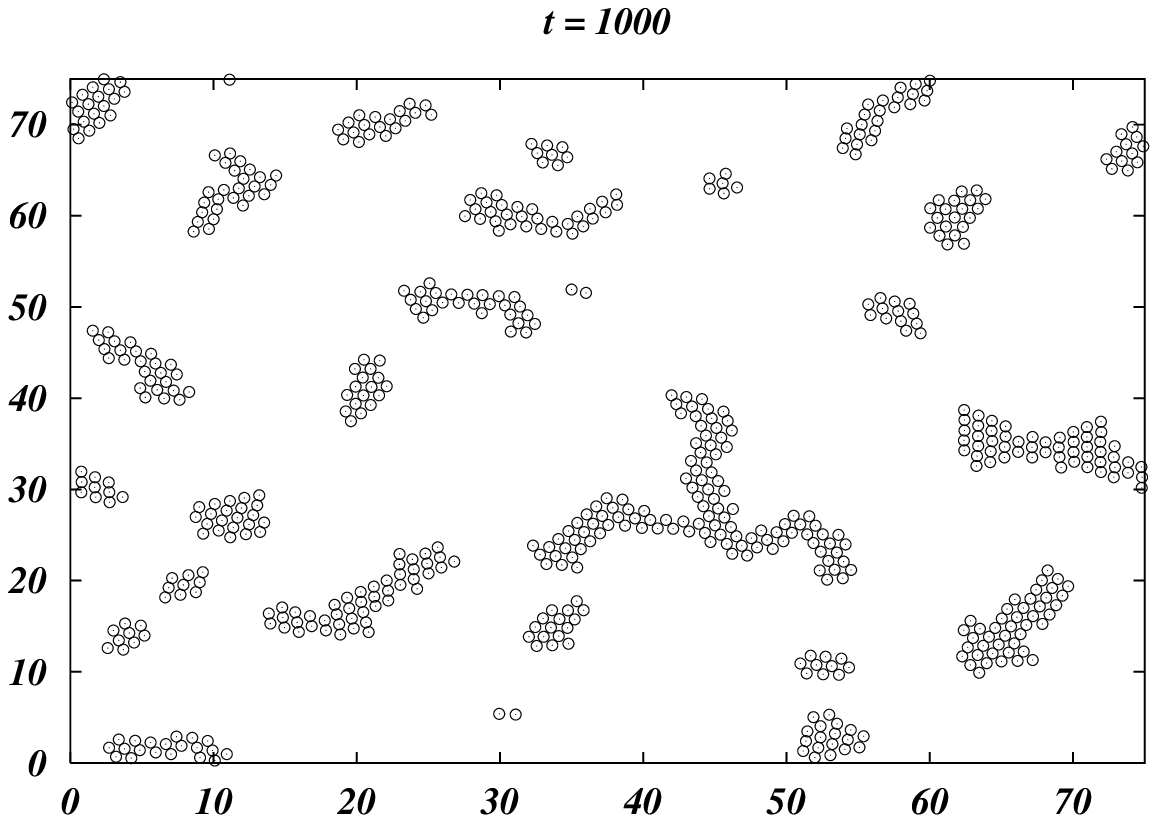,width=6.8cm}
      \epsfig{file=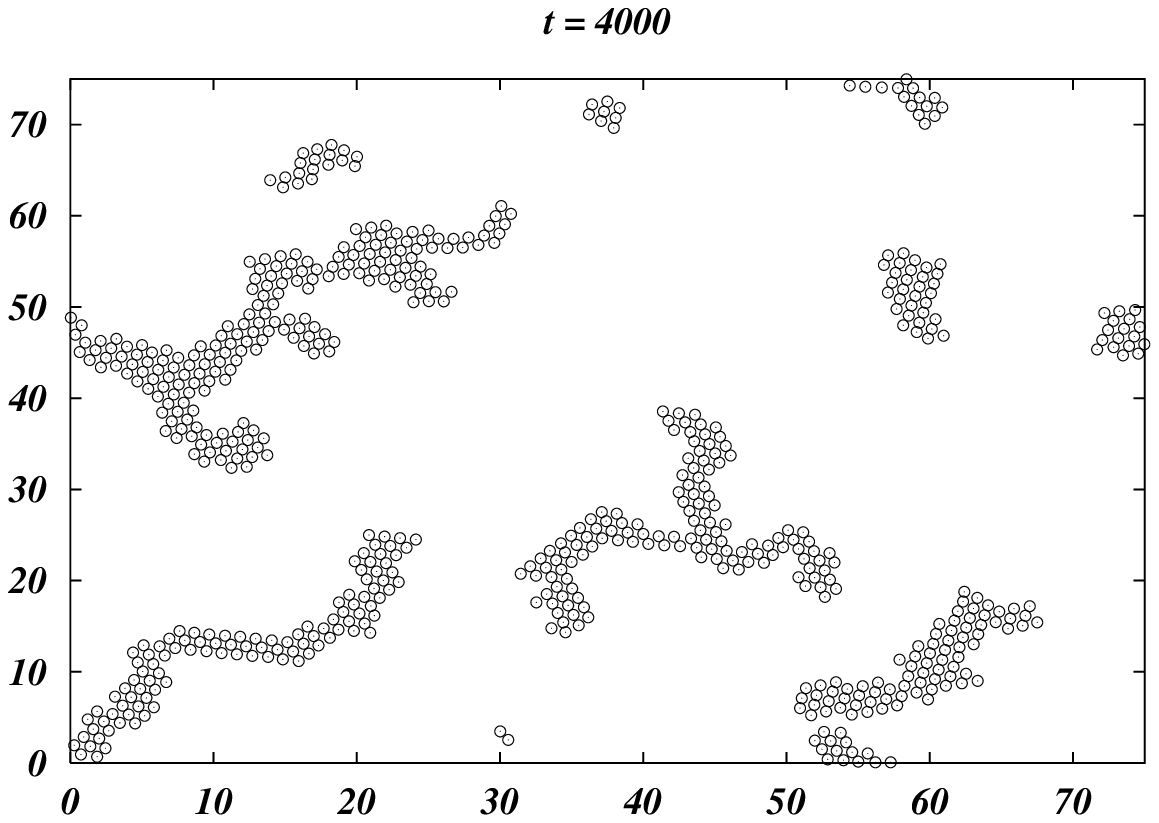,width=6.8cm}\\
    \end{center}
  \end{minipage}
  \begin{center}
    \caption{Time evolution of the cluster formation with 625 particles.}
    \label{fig:cells}
  \end{center}
\end{figure}

The collective motion of more complex $N$-particle systems (swarms) was
studied already in a series of papers by Vicsek et al. based on a spin glass
model in the velocity space \cite{ViCzBeCoSh95,CziVi00}.  First studies of the
dynamics of harmonic swarms of driven Brownian particles were reported
recently \cite{SchwEbTi01,EbSchw01b}. The dynamics of clusters of active
particles with Morse interactions was studied first in \cite{ErEbAn00}. Here
we restrict ourselves to Morse forces, generalizing our findings reported in
the previous work \cite{ErEbAn00}. Let us consider first the case that all
particles form one cluster which is hold together by relatively long range
Morse forces. Approximating the sum of forces in a mean field approximation we
may represent the local field near to the center of the swarm
\begin{equation}
  \bm{R}(t) = \frac{1}{N}\sum \bm{r}_i(t)
\end{equation}
by an anharmonic oscillator potential centered around $\bm R$, the center of
mass.  In the normal representation the potential reads
\begin{equation}
  U(x_1,x_2) = \frac{1}{2} \left[a_1 (x_1 - X_1) + a_2 (x_2 - X_2)\right]
\end{equation}
In arbitrary representations we have a second-rank tensor $\bm{a}$ with the
eigenvalues $a_1, a_2$, which determine the normal modes of oscillations
around the center. The tensor depends on the parameters of the force and on
the shape of the swarm. In this linear approximation the Langevin equation
for an individual Brownian particle in a $N$-particle system bears some
similarity to the parabolic case studied earlier \cite{SchwEbTi01,EbSchw01b}:
\begin{equation}
  \label{langev-ww}
  \dot{\bm{r}_i}=\bm{v_i}\quad ,\quad 
  m\dot{\bm{v}_i} + \bm{a}(\bm{r}_i - \bm{R}(t)) =
  - m\gamma(\bm{v}_i^2) \bm{v}_i - \sigma^2 (\bm v_i - V) 
   + \sqrt{2D} \bm{\xi}_i(t)  
\end{equation}
The difference to the case studied before is the anharmonicity and the
existence of a weak global coupling.  Neglecting first the effects of
anharmonicity we may predict, on the basis of the previous findings, that
clusters with rotational symmetry will show a translational and a rotational
mode very similar to the case $N = 2$. Therefore we expect to see translating
as well as rotating clusters which might change their sense of rotation due to
stochastic effects (see Fig.~\ref{fig:rot}). With increasing asymmetry of the
shape, rotating clusters are expected to get unstable similar as asymmetric
driven oscillators \cite{ErEbAn00}. In order to check these predictions at
least qualitatively we have carried out several simulations for swarms with
Morse interactions (see Figs. \ref{fig:rot} and \ref{fig:cells}).

The basic results of our observations may be summarized as follows:
\begin{description}
\item[cluster drift] We see in the simulations drifting clusters
 rotating very slowly and clusters without rotations which move rather fast.
 This corresponds to the translational mode studied for $N=2$. Here most of the
 energy is concentrated in the kinetic energy of translational movement.
\item[generation of rotations] As we see from the simulations, small Morse
  clusters up to $N \simeq 20$ generate left/right rotations around their
  center of mass. The angular momentum distribution is bistable. This
  corresponds to the rotational mode studied above for $N = 1,2$. 
\item[breakdown of rotations] The rotation of clusters may come to a stop due
  to several reasons.  The first is the anharmonicity of clusters. As we have 
  shown above \cite{ErEbAn00}, {\em strong anharmonicity 
     destroys} the rotational mode.  Another reason are noise induced  
  transitions, this will be investigated in a forthcoming  
  paper\cite{ErEbMi01}. 
\item[shape distribution] With increasing noise the
  shape of the clusters is amoeba-like and is getting more and more
  complicated. A theoretical interpretations of the shape dynamics is still
  missing. 
\item[cluster composition] With increasing noise we observe a distribution of
  clusters of different size. Again a theory of clustering in the
  two-dimensional case is still missing. For the case of one-dimensional rings
  with Morse interactions several theoretical results are available
  \cite{DuEbEr01,DuEbErMa01}.
\end{description}

\section{Conclusions}

We studied here the active Brownian dynamics of a finite number of
confined or self-confined particles with velocity-dependent friction.
Confinement was created 
\begin{itemize}
\item[(i)] by external parabolic forces, 
\item[(ii)] by attracting Morse interactions.
\end{itemize}
We have given here first an analysis of several simple cases as
the motion of noninteracting driven particles in external potentials 
and the collective and relative motion of two driven particles.
Based on these analytical investigations we have made several predictions for
the behavior of Morse clusters with small particle numbers. This way we could
identify several qualitative modes of movement. Further we have made a
numerical study of special $N$ particle Morse systems. In particular we
investigated the rotational and translational modes of the swarm and the
clustering phenomena.

We did not intend here to model any particular problem
of biological or social collective movement. We note however that the study of
dynamic modes of collective movement of swarms may be of some importance for
the understanding of many biological and social collective motions.
To support this view we refer to the book of Okubo and Levin
\cite{OkLe02} where the modes of collective motions of swarms
of animals are classified in a way which reminds very much the 
dynamical modes of the model investigated here.

\section*{Acknowledgments}
The authors are grateful to Yu. L. Klimontovich (Moscow), A. S. Mikhailov
(Berlin), F. Schweitzer (Birlinghoven) and L. Schimansky-Geier (Berlin) for
discussions.

\bibliographystyle{unsrt}
\bibliography{allgemein,bakterien,cells,coherent,erdmann,gbt,brown,toda,neu}

\begin{thebibliography}{10}

\bibitem{ViCzBeCoSh95}
Tam{\'a}s Vicsek, Andr{\'a}s Czir{\'o}k, Eshel Ben-Jacob, Inon Cohen, and
  O.~Shochet.
\newblock Novel type of phase transition in a system of self-driven particles.
\newblock {\em Physical Review Letters}, 75(6):1226--1229, August 1995.

\bibitem{DeVi95}
Imre Der{\'e}nyi and Tam{\'a}s Vicsek.
\newblock Cooperative transport of {B}rownian particles.
\newblock {\em Physical Review Letters}, 75(3):374--377, 1995.

\bibitem{ShiSuMiHaSa96}
Naohiko Shimoyama, Ken Sugawara, Tsuyoshi Mizuguchi, Yoshinori Hayakawa, and
  Masaki Sano.
\newblock Collective motion in a system of motile elements.
\newblock {\em Physical Review Letters}, 76(20):3870--3873, May 1996.

\bibitem{CziVi00}
Andr{\'a}s Czir{\'o}k and Tam{\'a}s Vicsek.
\newblock Collective behavior of interacting self-propelled particles.
\newblock {\em Physica A}, 281(1-4):17--29, 2000.

\bibitem{Ra45}
J.~W.~Strutt Rayleigh.
\newblock {\em The Theory of Sound}, volume~I.
\newblock Dover, New York, 2 edition, 1945.

\bibitem{Kl95}
Yu.~L. Klimontovich.
\newblock {\em Statistical Theory of Open Systems}.
\newblock Kluwer Acad. Publ., Dordrecht, 1995.

\bibitem{MiZa99}
Alexander~S. Mikhailov and Dami{\'a}n Zanette.
\newblock Noice-induced breakdown of coherent collective motion in swarms.
\newblock {\em Physical Review E}, 60(4):4571--4575, 1999.

\bibitem{SchiGr93}
M.~Schienbein and Hans Gruler.
\newblock Langevin equation, {F}okker-planck equation and cell migration.
\newblock {\em Bull. Mathem. Biology}, 55:585--608, 1993.

\bibitem{SchwEbTi98}
Frank Schweitzer, Werner Ebeling, and Benno Tilch.
\newblock Complex motion of {B}rownian particles with energy depots.
\newblock {\em Physical Review Letters}, 80(23):5044--5047, June 1998.

\bibitem{EbSchwTi99}
Werner Ebeling, Frank Schweitzer, and Benno Tilch.
\newblock Active {B}rownian particles with energy depots modelling animal
  mobility.
\newblock {\em BioSystems}, 49:17--29, 1999.

\bibitem{ErEbSchiSchw99}
Udo Erdmann, Werner Ebeling, Frank Schweitzer, and Lutz Schimansky-Geier.
\newblock Brownian particles far from equilibrium.
\newblock {\em European Physical Journal B}, 15(1):105--113, 2000.

\bibitem{MaEbVe99}
V.~Makarov, Werner Ebeling, and M.~Velarde.
\newblock Soliton-like waves on dissipative toda lattices.
\newblock {\em International Journal of Bifurcation and Chaos},
  10(5):1075--1089, 2000.

\bibitem{EbErDuJe99}
Werner Ebeling, Udo Erdmann, J\"orn Dunkel, and Martin Jenssen.
\newblock Nonlinear dynamics and fluctuations of dissipative toda chains.
\newblock {\em Journal of Statistical Physics}, 101(1/2):443--457, 2000.

\bibitem{Gr73}
R.~Graham.
\newblock Statistical theory of instabilities in stationary non-equilibrium
  systems with applications to lasers and nonlinear optics.
\newblock {\em Springer Tracts in Modern Physics}, 66:111, 1973.

\bibitem{Gr81}
R.~Graham.
\newblock Models of stochastic behaviour in non-equilibrium steady states.
\newblock In S.H. Chen, B.~Chu, and R.~Nossal, editors, {\em Scattering
  Techniques applied to supramolecular and non-equlibirum systems}, New York,
  1981. Plenum Press.

\bibitem{Ha73}
H.~Haken.
\newblock {\em Z. Phys.}, 263:267, 1973.

\bibitem{HoRy78}
M.O. Hongler and D.M. Ryter.
\newblock {\em Z. Phys. B}, 31:333, 1978.

\bibitem{Eb00}
Werner Ebeling.
\newblock Canonical nonequilibrium statistics and applications to fermi-bose
  systems.
\newblock {\em Condensed Matter Physics}, 3(2(22)):285--293, 2000.

\bibitem{TiSchwEb99}
Benno Tilch, Frank Schweitzer, and Werner Ebeling.
\newblock Directed motion of {B}rownian particles with internal energy depot.
\newblock {\em Physica A}, 273(3-4):294--314, 1999.

\bibitem{SchwEbTi01}
Frank Schweitzer, Werner Ebeling, and Benno Tilch.
\newblock Statistical mechanics of canonical-dissipative systems and
  application to swarm dynamics.
\newblock {\em Physical Review E}, 64:021110, 2001.

\bibitem{EbErSchiSchw99}
Werner Ebeling, Udo Erdmann, Lutz Schimansky-Geier, and Frank Schweitzer.
\newblock Complex motion of brownian particles with energy supply.
\newblock In David~S. Broomhead, Elena~A. Luchinskaya, Peter V.~E. McClintock,
  and Tom Mulin, editors, {\em Stochastic and Chaotic Dynamics in the Lakes},
  volume 502 of {\em AIP Conference Proceedings}, pages 183--190, Melville, New
  York, 2000. American Institute of Physics.

\bibitem{Eb01}
Werner Ebeling.
\newblock Nonequilibrium statistical mechanics of swarms of driven particles.
\newblock {\em Physica A}, 314:92--96, 2002.

\bibitem{EbLaUs01}
Werner Ebeling, Polina~S. Landa, and Vadim~S. Ushakov.
\newblock Self-oscillations in ring toda chains with negativ friction.
\newblock {\em Physical Review E}, 63:046601, 2001.

\bibitem{DuEbEr01}
J\"orn Dunkel, Werner Ebeling, and Udo Erdmann.
\newblock Thermodynamics and transport in an active morse ring chain.
\newblock {\em European Physical Journal B}, 24(4):511--524, 2001.

\bibitem{DuEbErMa01}
J\"orn Dunkel, Werner Ebeling, Udo Erdmann, and Valeri~A. Makarov.
\newblock Coherent motions and clusters in a dissipative morse ring chain.
\newblock {\em International Journal of Bifurcation and Chaos},
  12(11):2359--2377, November 2002.

\bibitem{ErEbAn00}
Udo Erdmann, Werner Ebeling, and Vadim~S. Anishchenko.
\newblock Excitation of rotational modes in two-dimensional systems of driven
  brownian particles.
\newblock {\em Physical Review E}, 65:061106, June 2002.

\bibitem{EbSchw01b}
Werner Ebeling and Frank Schweitzer.
\newblock Active motion in systems with energy supply.
\newblock In M.~Matthies, Horst Malchow, and J.~Kriz, editors, {\em Integrative
  Systems Approaches to Natural and Social Dynamics}, pages 119--142, Berlin,
  2001. Springer.

\bibitem{ErEbMi01}
Udo Erdmann, Werner Ebeling, and Alexander~S. Mikhailov.
\newblock Translational modes and noise induced transitions of swarms of active
  brownian particles.
\newblock Preprint, October 2002.

\bibitem{OkLe02}
Akira Okubo and Levin~Simon A.
\newblock {\em Diffusion and Ecological Problems: Modern Perspectives},
  volume~14 of {\em Interdisciplinary Applied Mathematics}.
\newblock Springer, New York, 2. edition, 2001.

\end{thebibliography}

\end{document}